\documentclass{iaus}
\usepackage{graphicx}

\title[IAU~285.~~New Horizons in Time Domain Astronomy] 
{Search for Electromagnetic Counterparts to LIGO-Virgo Candidates:
Expanded Very Large Array\thanks{The National Radio Astronomy Observatory is a facility of the National Science Foundation operated under cooperative agreement by Associated Universities, Inc.} Observations}

\author[Lazio et al.]
{Joseph Lazio${}^1$\thanks{Part of this research was carried
out at the Jet Propulsion Laboratory, California Institute of
Technology, under a contract with the National Aeronautics and Space Administration.},
Katie Keating${}^2$\thanks{A portion of this work was performed while K.~K.\ held a National Research Council-NRL Research Associateship.},
F.~A.~Jenet${}^3$\thanks{This work was supported by a CAREER grant from the National Science Foundation (award AST 0545837).},
N.~E.~Kassim${}^4$\thanks{Basic research in radio astronomy at the NRL
is supported by 6.1 Base funding.},
the LIGO Scientific Collaboration, 
\and
the Virgo Collaboration}

\affiliation{%
$^1$Jet Propulsion Laboratory, California Institute of
	Technology, 4800 Oak Grove Dr., M/S~138-308, Pasadena, CA
	91109, USA \\
	email: {\tt Joseph.Lazio@jpl.nasa.gov} \\[\affilskip]
$^2$National Research Council, Naval Research Laboratory, 4555
	Overlook Avenue~\hbox{SW}, Washington, DC 20375-5351 USA \\
	[\affilskip]
${}^3$Center for Gravitational Wave Astronomy, University of Texas at
	Brownsville, Brownsville, TX 78520, USA \\ 
	[\affilskip]
${}^4$Remote Sensing Division, Naval Research Laboratory, 4555
	Overlook Avenue~\hbox{SW}, Washington, DC 20375-5351 USA \\
	[\affilskip]
}

\pubyear{2011}
\volume{285}  
\pagerange{1--4}
\jname{New Horizons in Time Domain Astronomy}
\editors{R.~E.~M.~Griffin, R.~J.~Hanisch \& R.~Seaman}
\begin{document}

\maketitle

\begin{abstract}
This paper summarizes a search for radio wavelength counterparts to
candidate gravitational wave events.  The identification of an
electromagnetic counterpart could provide a more complete
understanding of a gravitational wave event, including such
characteristics as the location and the nature of the progenitor.  We
used the Expanded Very Large Array (EVLA) to search six galaxies which
were identified as potential hosts for two candidate gravitational
wave events.  We summarize our procedures and discuss preliminary results.
\end{abstract}

\firstsection 
\section{Gravitational Wave Astronomy and the Time Domain}\label{sec:jl.intro}

Gravitational waves (GWs) are fluctuations in the spacetime metric,
equivalent to electromagnetic waves resulting from fluctuations in an
electromagnetic field.  Because of the weakness of the gravitational
force, however, a laboratory demonstration of GWs comparable to
Hertz's demonstration of electromagnetic waves is not possible.
Indeed, the characteristic scale for the luminosity of a GW source is
$L_0 = 2 \times 10^5\,M_\odot c^2$~s${}^{-1}$, indicating immediately
that the generation of GWs will occur in astrophysical environments in
which large masses are moving at high velocities.

Precise timing of pulses from the radio pulsar PSR~B1913$+$16, which
is one member of a double neutron star system, has revealed indirect
evidence for GWs (\cite[Hulse \& Taylor~1975]{ht75}; \cite[Weisberg et
al.~2010]{wnt10}).  In this system, the rate of orbital period decay
as predicted by general relativity is consistent with that inferred
from the pulsar timing measurements.  Pulsars in other neutron star-neutron star binaries have since
been discovered, and the general relativistic rates of orbital period decay remain consistent with those measured.

From the standpoint of time domain astronomy, many of the predicted
sources of GWs are rapidly time varying phenomena, and Centrella et
al.~(2011) summarize one of the Symposium workshops.  (See also
\cite[Sathyaprakash \& Schutz~2009]{ss09}.)  Potential sources include
the mergers of compact objects (neutron star-neutron star, neutron
star-black hole, and black hole-black hole mergers), asymmetric
supernovae, rapidly rotating asymmetric neutron stars, and exotic
objects such as oscillating cosmic strings.

While evidence for GWs remains indirect, the promise of direct
detection has excited considerable international interest.  \textit{A Science Vision for European Astronomy}
posed ``Can we observe strong gravity in action?'' as a key question
for this decade while, in the U.{}S., {GW astronomy}
was identified as a scientific frontier discovery area in the
\textit{New Worlds, New Horizons in Astronomy \& Astrophysics}
Decadal Survey.  GW astronomy is expected to be similar to the
experience in opening up new spectral windows in the electromagnetic
spectrum.  As each new electromagnetic spectral window has been
opened, entirely new classes of sources have been discovered.
Indeed, one of the most surprising results of GW astronomy would be if
\emph{no} new classes of sources were discovered.

\section{Electromagnetic Counterparts to Gravitational Wave
	Events}\label{sec:jl.counterpart}

Many GW source classes are expected to display
electromagnetic counterparts.  The benefits of identifying an electromagnetic
counterpart include
\begin{itemize}
\item Precise localization of the event, which may be crucial in
understanding the nature of the event (e.g., in the nucleus of a galaxy vs.\ at its outskirts).
\item The characteristics of the counterpart, such as its spectrum,
will likely constrain the environment or progenitor or both of the GW event.
\end{itemize}
In many respects, determining the electromagnetic counterpart to a GW
event is analogous to determining the (electromagnetic) spectrum of a
transient discovered in one band and then followed up at others (e.g.,
a gamma-ray burst followed up at X-ray, optical, and radio wavelengths).

Our focus is on radio counterparts, motivated by several
considerations (see also \cite[Predoi et al.~2010]{pcc+10}):
\begin{itemize}
\item Non-thermal, high energy particles often emit at radio
wavelengths, particularly in the presence of a magnetic
field (e.g., cyclotron and synchrotron emission).
\item Precise astrometry can be obtained, in the best
cases at the milliarcsecond level.
\item Dust obscuration, either
from the immediate environment of the event or from intervening
objects, is not an issue.
\item Radio telescopes can observe during the day, offering rapid
followup.
\item If a GW event produces a radio burst or pulse, 
the ionized interstellar (and intergalactic) medium will delay the 
propagation of the pulse.  Such delays can be minutes to hours,
depending upon the electron column density along the line of sight,
but they potentially allow for detailed followup of the burst.
\end{itemize}

\section{LIGO-Virgo Observations}\label{sec:jl.gw}

The radio observations that we describe below are based on coordinated
observations between the Laser Interferometric Gravitational-wave
Observatory (LIGO) and Virgo that occurred during the Autumn of~2010
(\cite[LIGO Scientific Collaboration \& Virgo
Collaboration~2011]{l-v11}).  LIGO has two elements (\cite[Abbott et
al.~2009]{aaa+09}), located in Hanford, \hbox{WA}, \hbox{USA}, and
Livingston, \hbox{LA}, \hbox{USA}, while Virgo has one element located
near Pisa, Italy (\cite[Acernese et al.~2008]{aaa+08}).  Together they
form a 3-element interferometer.

The LIGO-Virgo interferometer measures time differences of arrival.
Analysis of test waveforms injected into the LIGO-Virgo processing
pipeline indicate that a candidate's position can be localized only
to tens of square degrees.  
Ordinarily, such a large region could not usefully be
searched for a radio counterpart with the current
generation of telescopes, because their fields of view are too small.
However, the most likely GW sources that 
could be detected at reasonable signal-to-noise ratios would have occurred
within~50~Mpc.  The galaxies within this horizon can be ranked
(\cite[Nuttall \& Sutton~2010]{ns10}; \cite[White et
al.~2010]{wdd11}), and the three most probable galaxies were selected
for observations.

\section{Expanded Very Large Array (EVLA) Observations}\label{sec:jl.evla}

The EVLA is a 27-element radio
interferometer operating between~1 and~50~GHz.  It has been the focus
of a recent major upgrade, which is nearly complete, and it is
being commissioned with science programs now well established.
For the purpose of radio followup of GW events, the EVLA offers a number of
attractive features.
\begin{itemize}
\item The wide frequency (wavelength) coverage potentially allows
``tuning'' of the observations to  match the
expected physics.  In this case, we observed at~5~GHz ($\lambda6$cm),
at which both expanding synchrotron fireballs and relativistic jets
are likely to be detectable.

\item At our observational frequency, the nominal field of view is
approximately $7^\prime$, well matched to the size of most local
galaxies.  In practice, the field of view is usually defined as the
region over which the antenna response is at least half of its peak
value; sources outside of the nominal field of view can still be
detected, provided that they are sufficiently strong to compensate for
the decreased antenna response.
Accordingly, in order not to miss a potential counterpart, we imaged a
much larger region ($\approx 30^\prime$).

\item The angular resolution of the EVLA can be adjusted by moving the
individual antennas.  During most of our observations, the obtained
angular resolution was about~$4^{\prime\prime}$, which provides
about~$0.\!\!^{\prime\prime}4$ localization ($\approx 20$~pc
at~10~Mpc), for reasonable signal-to-noise ratios.
\end{itemize}

\begin{figure}[tb]
\begin{center}
 \includegraphics[width=0.47\textwidth]{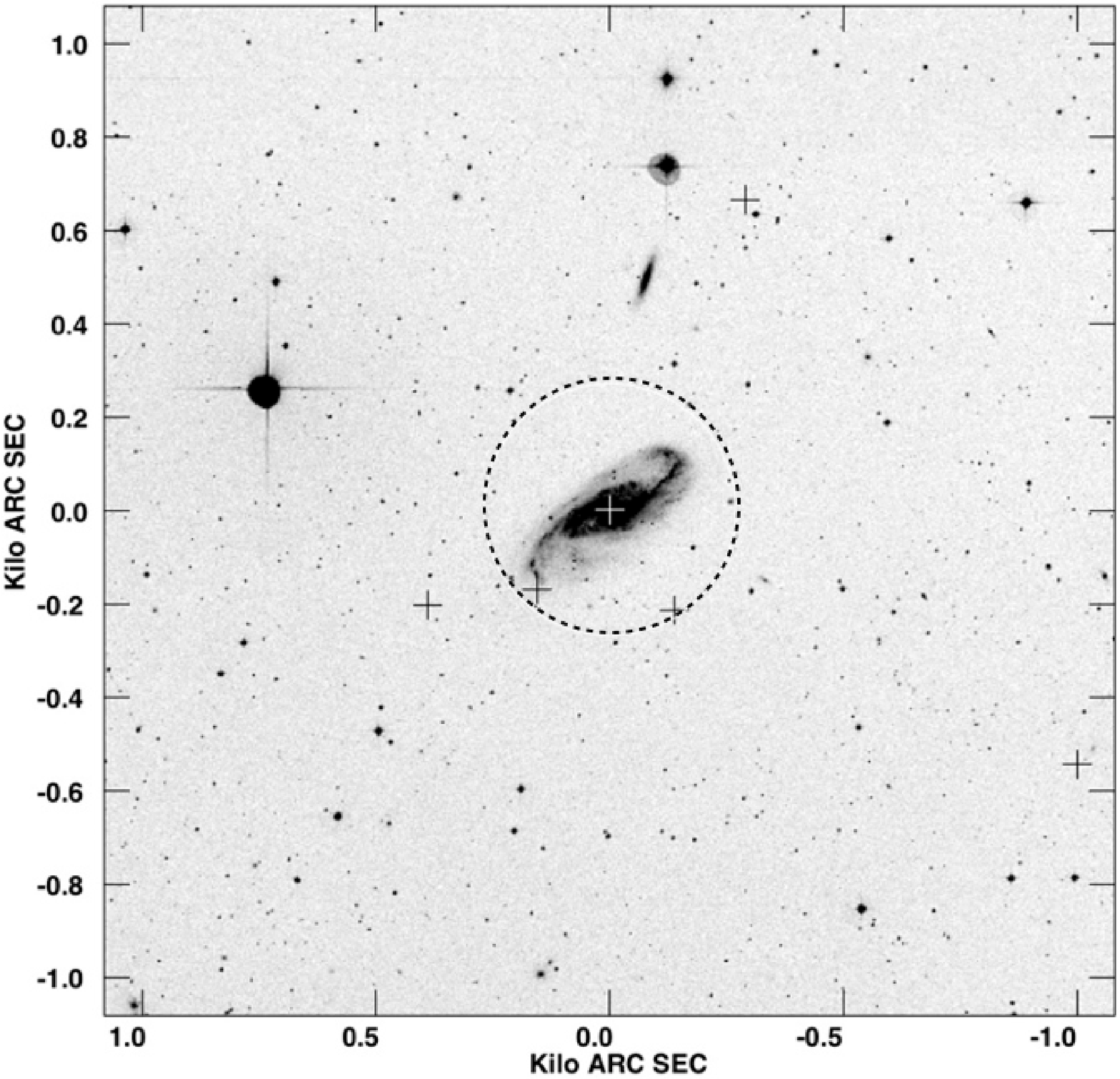}\,\includegraphics[width=0.47\textwidth]{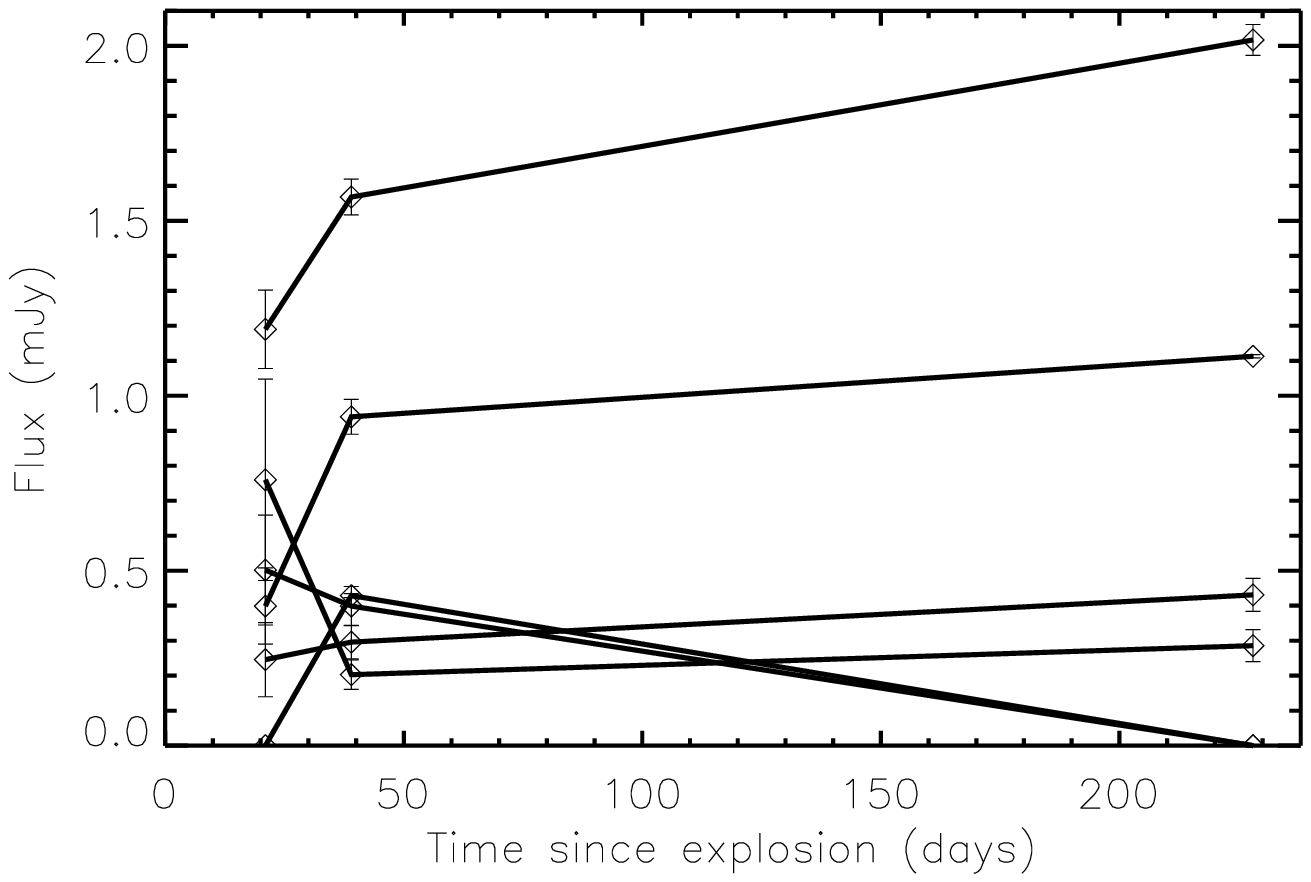}
\vspace*{-2ex}
 \caption{Radio wavelength ($\lambda6$cm) sources detected in the field of one of the galaxies representing a potential host for a
candidate LIGO-Virgo gravitational wave event and the accompanying
light curves.  Crosses indicate the
detected sources, and the circle shows where the EVLA antenna
response has decreased by a factor of two.}
\label{fig:jl.evla}
\end{center}
\end{figure}

We have conducted three epochs of observations for each of the two
LIGO-Virgo candidates, observing the three most probable hosts within the
uncertainty region for each candidate.
Figure~\ref{fig:jl.evla} shows the field
around one of the galaxies.  We typically detected six sources in the
field of each galaxy.  This small number of sources is consistent with the
number of extragalactic sources expected (\cite[{Windhorst}~2003]{w03}) but does not
exclude the possibility that one of the radio sources is a counterpart to
the LIGO-Virgo candidate.

Data acquisition and reduction of all three epochs for both candidates
has only recently been concluded.  In assessing the reality of any
potential radio counterpart to either LIGO-Virgo candidate,
other potential sources of variability must also be considered.  On
the time scales and cadence of our observations, it is unlikely that
intrinsic variability of any active galactic nuclei (AGN) in the field
of view would be a source of contamination.  However, refractive interstellar scintillation due to the Galaxy's
interstellar medium is a potential source of contamination (i.e., an
AGN unrelated to the LIGO-Virgo candidate might show variability on
the time scales of our observations).

\section{Future}\label{sec:jl.future}

Both LIGO and Virgo are currently being upgraded (\cite[Abadie et
al.~2010]{aaa+10}).  When they
resume observation ($\sim 2014$), they will be more sensitive and able
to probe to larger distances.
This increased GW sensitivity will require a change in observing
strategy.  A much larger number of galaxies could now be hosts, and a
search of all of them with the \hbox{EVLA} would likely be quite time
consuming (see also \cite[{Metzger \& Berger}~2011]{mb11}).  An
approach similar to current followup of supernovae and gamma-ray
bursts could, however, be profitable, namely, if a counterpart is
found at another wavelength, then the EVLA could assess its radio
properties.

Later in the decade, a number of other radio wavelength facilities are
likely to be available and present additional opportunities.
Among these are the Low Frequency Array (\hbox{LOFAR}, \cite[Fender~2011]{f11}), which
could conduct wide-field of view ``blind'' searches at meter
wavelengths (30--240~MHz) for northern hemisphere counterparts; the
Karoo Array Telescope (MeerKAT), which could conduct southern
hemisphere observations similar to those of the \hbox{EVLA};
and the Australian Square Kilometre Array Pathfinder (ASKAP), which
could conduct wide-field of view ``blind'' searches at decimeter
wavelengths ($\sim 1$~GHz) for southern hemisphere counterparts.  In
fact, LOFAR has acquired initial observations, similar to our EVLA
observations; those data are also under analysis.


%
%
%


\begin{thebibliography}{}
\bibitem[Abadie et al. (2010)]{aaa+10}
	Abadie, J., Abbott, B.~P., Abbott, R., et al.  2010,
	\textit{Classical Quant.\ Grav.}, 27, 173001

\bibitem[Abbott et al. (2009)]{aaa+09}
	Abbott, B.~P., Abbott, R., Adhikari, R., et al.  2009, 
	\textit{Rep.\ Prog.\ Phys.}, 72, 076901

\bibitem[Acernese et al. (2008)]{aaa+08}
	Acernese, F., Alshourbagy, M., Amico, P., et al.  2008, 
	\textit{Classical Quant.\ Grav.}, 25, 184001

\bibitem[Centrella et al. (2011)]{cnw11}
	Centrella, J., Nissanke, S., \& Williams, R.  2011, in New
	Horizons in Time Domain Astronomy, eds.\ R.~E.~M.~Griffin,
	R.~J.~Hanisch \& R.~Seaman, this volume

\bibitem[Fender (2011)]{f11}
	Fender, R.  2011, in New
	Horizons in Time Domain Astronomy, eds.\ R.~E.~M.~Griffin,
	R.~J.~Hanisch \& R.~Seaman, this volume

\bibitem[{Hulse \& Taylor} (1975)]{ht75}
	{Hulse, R.~A., \& Taylor, J.~H.} 1975,
	\textit{ApJ}, 195, L51

\bibitem[LIGO Scientific Collaboration \& Virgo Collaboration
(2011)]{l-v11}
	LIGO Scientific Collaboration \& Virgo Collaboration, 2011,
	arXiv:1109.3498

\bibitem[{Metzger \& Berger} (2011)]{mb11}
	{Metzger, B.~D., \& Berger, E.}  2011, 
	\textit{ApJ}, in press; arXiv:1108.6056

\bibitem[Nuttall \& Sutton (2010)]{ns10}
	Nuttall, L.~K., \& Sutton, P.~J.  2010,
	\textit{PRD}, 82, 102002

\bibitem[Predoi et al. (2010)]{pcc+10}
	Predoi, V., Clark, J., Creighton, T., et al.  2010, 
	\textit{Classical Quant.\ Grav.}, 27, 084018

\bibitem[Sathyaprakash \& Schutz (2009)]{ss09}
	Sathyaprakash, B.~S., \& Schutz, B.~F.  2009, 
	\textit{Living Rev.\ Relativ.}, 12, 2

\bibitem[{Weisberg \etal} (2010)]{wnt10}
	{Weisberg, J.~M., Nice, D.~J., \& Taylor, J.~H.}  2010, 
	\textit{ApJ}, {722}, 1030

\bibitem[White et al. (2011)]{wdd11}
	White, D.~J., Daw, E.~J., \& Dhillon, V.~S.  2011,
	\textit{Classical Quant.\ Grav.}, 28, 085016

\bibitem[{Windhorst} (2003)]{w03}
	{Windhorst, R.~A.}  2003, 
	\textit{New Astron.\ Rev.}, 47, 357

%
%
%
%
%
%
%
%
%
%
%
%
%
%
%
%
%
\end{thebibliography}
\end{document}